\pdfoutput=1 
\documentclass[10pt,conference]{IEEEtran}
\usepackage{lipsum,multicol}
\usepackage{color}

\newtheorem{theorem}{Theorem}[section]
\newtheorem{lemma}[theorem]{Lemma}
\newtheorem{proposition}[theorem]{Proposition}

\newtheorem{defn}[theorem]{Definition}
\newtheorem{remark}{Remark}

\newenvironment{example}[1][Example]{\begin{trivlist}
\item[\hskip \labelsep {\bfseries #1}]}{\end{trivlist}}
\usepackage{graphics} %
\usepackage{cite}
\usepackage{color}
\usepackage{float}
\usepackage{algorithmic}
\usepackage{amsmath}
\usepackage{amsfonts}
\usepackage{amssymb}
\usepackage{amsopn}
\usepackage{xspace}
\usepackage{array}
\usepackage{epsfig}
\usepackage{color}
\usepackage{subfigure}
\usepackage{multirow}

\newcommand{\Ex}{\mathop{\bf E\/}}

\newcommand{\modH}{\hat{\mathbf{H}}}

\newcommand{\modV}{\hat{\mathbf{V}}}

\newcommand{\rankH}{r}

\newcommand{\rate}{R}

\newcommand{\barZi}{|\mathcal{S}_{i}|}

\makeatletter  
 \renewcommand\subsubsection{\@startsection{subsubsection}{3}{\z@}%
                        {-18\p@ \@plus -4\p@ \@minus -4\p@}%
                        {8\p@ \@plus 4\p@ \@minus 4\p@}
                        {\normalfont\normalsize\bfseries\boldmath
                         \rightskip=\z@ \@plus 8em
                          \pretolerance=10000 }}
                          \makeatother   

\begin{document}

\title{Delay Constrained Throughput Analysis of a Correlated MIMO Wireless Channel}

%

\author{\IEEEauthorblockN{Kashif Mahmood\IEEEauthorrefmark{1}, Mikko Vehkaper{\"a}\IEEEauthorrefmark{2}, Yuming Jiang\IEEEauthorrefmark{1}}       

\IEEEauthorblockA{\IEEEauthorrefmark{1} Q2S, Norwegian University of Science and Technology,~Norway\\
\IEEEauthorrefmark{2} School of Electrical Engineering and the ACCESS Linnaeus Center,~KTH,~Sweden  }}


\maketitle

\begin{abstract}
The maximum traffic arrival rate at the network for a given delay guarantee (\emph{delay constrained throughput}) has been well studied for wired channels.
However, few results are available for wireless channels, especially when multiple antennas are employed at the transmitter and receiver.
In this work, we analyze the network delay constrained throughput of a multiple input multiple output (MIMO) wireless channel with time-varying spatial correlation.
The MIMO channel is modeled via its virtual representation, where the
individual spatial paths between the antenna pairs are Gilbert-Elliot channels.  The whole system is then described by a $K$-State Markov chain, where $K$ depends upon the degree of freedom (DOF) of the channel.
We prove that the DOF based modeling is indeed accurate.
Furthermore, we study the impact of the delay requirements at the network layer, violation probability and the number of antennas on the throughput under different fading speeds and signal strength.
\end{abstract}


\section{Introduction}
Multiple input multiple output (MIMO)~\cite{MIMO:Foschini:98:limits}~\cite{MIMO:Teletar99:CapacityOfMultiAntennaGaus} communication has generated a lot of interest in recent years for providing high data rates and reliability without consuming extra power or bandwidth. It significantly improves the capacity as compared to traditional single input single output (SISO) communication.
The importance of MIMO technology can be well understood from the fact that IEEE standardization committee is actively pursuing IEEE $802.11$n and $802.16$e which promise high throughput using multiple antennas in  wireless networks.

Significant amount of work has been done on MIMO to predict the rate regions under different channel conditions, antenna settings, etc.,
under the assumption of infinite delay and queue / buffer sizes \cite{MIMO:Teletar99:CapacityOfMultiAntennaGaus}.
Recently, however, there has been a rapid growth of delay sensitive applications like VoIP and IP video that are  run on packet switched wireless networks.
Albeit the decoding delay at the physical layer has been reduced by the invention of graph-like codes to the order of thousands of bits, the delay%
\footnote{Throughout the paper, the term \emph{delay} refers to \emph{queuing delay} and the \emph{throughput} as the \emph{delay constrained throughput}.}
at the higher layers has a significant impact on the overall performance for such systems.
In addition, MIMO links have been considered for delay critical applications such as mobile ad hoc networks (MANETs)~\cite{MIMO:Adhoc:Chen06}, for they overcome the problems of low system throughput.
It is therefore essential to predict the rate region of a MIMO system for a given queuing delay guarantee.
Indeed, the task of predicting the rate regions for a given delay guarantee is seen as one of the grand challenges in wireless networks~\cite{NIT:Jan08:RethinkingIT}.

In this paper, we formulate a method to find the \emph{delay constrained throughput} of a spatial multiplexed \emph{correlated} MIMO wireless channel on the flow level.
We define the delay constrained throughput as the \emph{maximum packet arrival rate} at the network for which the delay requirement is met.
The service model is based on the Gilbert-Elliott (GE) channel~\cite{WirelessMarkov:Gilbert60:CapacityBurstNoiseChannel, elliot_noise} where for a given setting of the propagation environment, each channel path is considered in the \emph{virtual domain}  \cite{MIMO:Correlated:Sayeed02:Deconstructing,MIMO:Veeravalli2005:CorrelatedMIMO:Variance}  to be either in a $\mathsf{good}$ or a $\mathsf{bad}$ state.
The $\mathsf{good}$ represents the case when there is a strong wireless link between the antenna pairs, while the $\mathsf{bad}$ corresponds to a scenario when the link is blocked by some large obstacle(s).
The permutations of the Tx-Rx channel paths comprise the sub-states of the Markov chain describing the whole system.
The sub-states are aggregated based on the degrees of freedom (DOF) available of the MIMO channel~\cite{MIMO:QoS:FSM:Kashif} which results in a $K$-State Markov chain, where $K$ depends upon the DOF.
The DOF based modeling results in the reduction of the state space from exponential in the number of antennas to linear in the DOF.
In addition to numerical simulations as in \cite{MIMO:QoS:FSM:Kashif}, the method is justified here by a mathematical proof.

Multi-antenna systems typically operate in one of the following transmission modes.
\emph{Spatial multiplexing} is used to increase throughput by transmitting \emph{independent} data flows over different antennas.
\emph{Spatial diversity}, on the other hand, increases reliability by transmitting \emph{redundant} flows of information in parallel.
There has been some work reported for MIMO diversity systems~\cite{EBWireless:Tang2007:CrossLayerModelingforQoS,MIMO:Diversity:Zorzi99:Lateness}. In \cite{EBWireless:Tang2007:CrossLayerModelingforQoS}, the concept of effective capacity has been used to carry out QoS analysis of independent and identically distributed (i.i.d) MIMO systems using finite state Markov chains to describe the received signal to noise ratio (SNR).
However, it is well known~\cite{MIMO:Correlated:Sayeed02:Deconstructing,MIMO:Veeravalli2005:CorrelatedMIMO:Variance} that the i.i.d fading model is highly idealistic --- a realistic MIMO channel model is one in which the elements of the channel matrix are correlated.  Furthermore, exploiting the DOF provided by the MIMO for spatial multiplexing is far more important for high-speed wireless systems than spatial diversity considered in \cite{EBWireless:Tang2007:CrossLayerModelingforQoS,MIMO:Diversity:Zorzi99:Lateness}.

A significant amount of work has been done on link-layer channel modeling of wireless channels~\cite{WirelessMarkov:Sadeghi08:FiniteStateMarkov:Survey}.
In~\cite{Wireless:queuing:bisnik2009}, diffusion approximation is used to carry out the throughput analysis of multihop memoryless wireless networks under SISO communication.
The concept of effective bandwidth is used in \cite{EBWireless:Li07} to estimate the capacity of the wireless channels with memory that varies randomly with time and space.
However none of these works covered the throughput analysis of more realistic spatially multiplexed correlated MIMO systems.

The major contribution of this work is as follows.
We present a methodology to calculate the \emph{network delay constrained throughput} of a \emph{correlated} MIMO system that uses spatial multiplexing at the transmitter.
In the analysis, we incorporate statistical independence of arriving traffic and the service provided by the channel.
Secondly, we give a mathematical proof that the DOF based aggregation used in \cite{MIMO:QoS:FSM:Kashif} is indeed accurate.
Furthermore we show the impact of increasing the delay guarantee and the number of antennas on the throughput under various conditions, such as signal strength and fading speed.

The rest of the paper is organized as follows.
We first present the correlated MIMO system model in Sec.~\ref{sec:prelim}. A Markov modeling of the correlated MIMO system based on the DOF is carried out in Sec.~\ref{Sec:DelayConstrainedThAnalysis} for calculating the delay constrained throughput.
An application scenario based on IEEE $802.11$n is presented in Sec.~\ref{Sec:NumericalResults} for which we calculate the numerical results.

\section{The System}
\label{sec:prelim}
%
%

We consider a single user MIMO system with $N$ transmit and $M$ receive antennas and spatial multiplexing at the transmitter.
Throughout the paper, perfect channel knowledge is assumed to be available at the receiver while the transmitter has no information about the propagation environment.
We use $[\cdot]^\dag$ to denote the matrix conjugate transpose, $\mathbf{I}_n$ denotes the identity matrix of order $n$, while $\mathrm{Var}(\cdot)$ denotes the variance and $\mathcal{CN}(\mu,\psi^2)$ denotes the circularly symmetric complex Gaussian distribution with mean~$\mu$ and variance~$\psi^2$.
The $M \times 1$ received signal vector $\mathbf{y}$ of a MIMO system reads
\begin{equation}
	\label{eq:system_model}
\mathbf{y} = \sqrt{\frac{\rho}{N}} \mathbf{Hx}+\mathbf{n} \enspace ,	
\end{equation}
where $\mathbf{x}$ is a $N \times 1$ transmitted signal vector, $\mathbf{n} \sim \mathcal{CN}(\mathbf{0},\mathbf{I}_M)$ and $\rho$ is the average total SNR at one receiving antenna.
$\mathbf{H}$ is an $M \times N$ complex channel matrix with entries $h_{m,n}$, where a particular $h_{m,n}$ represents the channel path from transmit antenna $n$ to receive antenna $m$.
The channel gains are normalized so that $\Ex [|h_{m,n}|^2]=1,\forall m,n$.

In most of the literature on MIMO channels (see for example \cite{MIMO:Foschini:98:limits}~and~\cite{MIMO:Teletar99:CapacityOfMultiAntennaGaus}), the channel matrix~$\mathbf{H}$ is assumed to have i.i.d complex Gaussian entries.
The i.i.d  MIMO channel model represents the ideal case where the environment is rich scattering and antenna spacing sufficiently large.
It is known, however, that that this is seldom the case in practice \cite{MIMO:Correlated:Sayeed02:Deconstructing,MIMO:Veeravalli2005:CorrelatedMIMO:Variance}.
In this work, we use a more realistic model where the elements of the channel matrix are \emph{correlated} and the correlation evolves in time according to a discrete time Markov chain.

The virtual model proposed in \cite{MIMO:Correlated:Sayeed02:Deconstructing, MIMO:Veeravalli2005:CorrelatedMIMO:Variance} is a convenient method to link the physical propagation environment to the correlation structure of the MIMO channel when the correlation varies slowly compared to the Rayleigh fading experienced by the individual channel paths.  The true channel under the virtual model reads
\begin{equation}
\label{E:VMIMO}
\mathbf{H}=\mathbf{A}_N \widetilde{ \mathbf{H} }  \mathbf{A}^{\dag}_M   \enspace ,
\end{equation}
where $\widetilde{ \mathbf{H} } $ represents the virtual channel related to $\mathbf{H}$, while $\mathbf{A}_N$ and $\mathbf{A}_M$ are unitary discrete Fourier matrices
(cf.\ \cite{MIMO:Correlated:Sayeed02:Deconstructing,MIMO:Veeravalli2005:CorrelatedMIMO:Variance}).
Of the many useful properties of $\widetilde{ \mathbf{H} } $, the key observation is that the elements~$\widetilde{h}_{m,n}$ of $\widetilde{ \mathbf{H} } $ are \emph{independent} random variables when the number of antennas is sufficiently large.
Furthermore, the second-order statistics of $\widetilde{h}_{m,n}$ are directly related to the scattering function in \emph{virtual domain} and approximately given by the variance matrix~$\mathbf{V}$ with entries $v_{m,n}=\mathrm{Var}(\widetilde{h}_{m,n})$.
Since the original channel matrix $\mathbf{H}$ is normalized to have unit power entries, it follows from \eqref{E:VMIMO} that the variance matrix must satisfy
\begin{equation}\label{E:normalization_v}
\sum_{m,n}v_{m,n}=NM \enspace.
\end{equation}

It is to be highlighted that the virtual channel $\widetilde{ \mathbf{H} } $ with a variance matrix $\mathbf{V}$ generates a correlation structure on the true  channel~$\mathbf{H}$ as given in \cite[Eq.~(9)]{Raghavan:2010:WKM}.  This in turn maps to a specific physical scattering environment.
For simplicity, we let the scattering environment and hence the spatial correlation matrix, change between code words and consider the fading experienced by each of the code words to be ergodic.  This models a scenario where the positions of large objects (e.g.\ buildings) between the user terminal and the base station change relatively slowly, whereas the scattering environment around the user (induced e.g.\ by cars)  experiences rapid changes.
If we let the MIMO channel $\mathbf{H}$ be Rayleigh fading, the virtual channel matrix~$\mathbf{ \widetilde{H} }$ has (approximately) independent complex Gaussian entries $\widetilde{h}_{m,n} \sim \mathcal{CN}(0,v_{m,n})$.  Since we assume no CSI at the transmitter, the inputs are chosen to be independent standard complex Gaussian and the ergodic capacity conditioned on $\mathbf{V}$ reads \cite{MIMO:Veeravalli2005:CorrelatedMIMO:Variance}
\begin{equation}
	\label{E:MIMOCapFormulalogDetH}
\overline{C}(\mathbf{V}) =  \Ex \left[ \log_{2} \left [\mathrm{det} \left ( \mathbf{I}_M + \frac {\rho}{N} \mathbf{\widetilde{H}\widetilde{H}^{\dag}}   \right )     \right ] \right ] \enspace,
\end{equation}
where the expectation is with respect to all Gaussian channel matrices $\widetilde{\mathbf{H}}$ that have variance matrix $\mathbf{V}$.

%
%
\section{Delay Constrained Throughput Analysis}
    \label{Sec:DelayConstrainedThAnalysis}
The capacity \eqref{E:MIMOCapFormulalogDetH} depends on the scattering environment via matrix $\mathbf{V}$.  Instead of considering one specific $\mathbf{V}$ for all transmissions or picking it e.g.\ randomly, we let the propagation environment evolve between the code words according to a discrete time Markov chain with state space $\mathcal{S}$.  Note that although we focus in the following on the simplified setting where the number of antennas at both ends is $N$, generalization to arbitrary $M \times N$ MIMO channels is also possible. We shall omit these cases for the ease of exposition.

\subsection{Markov Modeling}
    \label{Sec:MarkovModleingCorMIMO}

%
%
The Markov chain describing the scattering environment is defined through the individual spatial paths by modeling them as identical and independent Gilbert-Elliott (GE) channels in the virtual domain.
We use the notation $S(\widetilde{h}_{m,n})$ to denote the state of the link $(m,n)$ and consider
for simplicity only two states, 
$\mathsf{good}$ ('$\mathsf{g}$') or $\mathsf{bad}$ ('$\mathsf{b}$'), so that
\begin{equation}\label{E:state_of_link_mn}
	S(\widetilde{h}_{m,n})\in\{\mathsf{b},\mathsf{g}\},\qquad \forall m,n = 1,\ldots,N \enspace.
\end{equation}
%
%
The \emph{sub-states}%
\footnote{We use the term \emph{sub-state} instead of \emph{state} in order to separate the state space of two different Markov chains.  Details are given below.}
$\mathbf{s}\in\mathcal{S}$ of the Markov chain describing the time evolution of the variance matrix $\mathbf{V}$ are then written as
\begin{equation}
	\mathbf{s} = \big[S(\widetilde{h}_{1,1}) \; \cdots \;\, S(\widetilde{h}_{N,N})\big] \in \{\mathsf{b},\mathsf{g} \}^{1\times N^{2}} \enspace.
\end{equation}
For the GE channels, the $\mathsf{good}$ state corresponds to a strong wireless link between the given antenna pairs (in the virtual domain), while a $\mathsf{bad}$ channel represents the case when the link is blocked by some large obstacle(s).
%
%
For simplicity, we follow the framework of \cite{Raghavan:2010:WKM} and let the $\mathsf{bad}$ state of the link $(m,n)$ be defined as $v_{m,n}=0$, while the $\mathsf{good}$ state corresponds to $v_{m,n}=v_{\mathbf{s}} \geq 1$.  Note that the variable $v_{\mathbf{s}}$ depends on the sub-state
$\mathbf{s}\in\mathcal{S}$ of the Markov chain
but not on the path indices, and is selected for all $\mathbf{s}\in\mathcal{S}$ so that \eqref{E:normalization_v} is satisfied.  Albeit this is the simplest possible scenario and selected for expository reasons, extensions to non-binary $\mathbf{V}$ are possible.

The probability for a particular link $(m,n)$ to go from the $\mathsf{good}$ state to the $\mathsf{bad}$ state is given by $p_{\mathsf{gb}}$ and from the $\mathsf{bad}$ state to the $\mathsf{good}$ state by $p_{\mathsf{bg}}$.  For a certain path, the steady state probability for the $\mathsf{bad}$ channel state reads thus
\begin{equation}
\kappa=\frac{p_{\mathsf{gb}} } { p_{\mathsf{gb}}+p_{\mathsf{bg}} }   .
\label{eq:block_error_prob}
\end{equation}
It is to be noted that the Markovian model of the time varying correlated MIMO channel is completely specified by the two independent parameters $\kappa$ and $p_{\mathsf{bg}}$ which describe the fading speed, i.e., the speed of change of the scattering environment. These parameters in turn
can be adjusted for example to meet the results of a measuring campaign.

\subsection{State Aggregation Based on DOF}
\label{subsec:StateAggBasedonDOF}

Since all individual links $(m,n)$ are independent two-state GE channels in the virtual domain, there is by definition a total of $|\mathcal{S}| = 2^{N^{2}}$ sub-states in the Markov chain describing the scattering environment.
We call this  \emph{method~$:1$}. The problem of this approach is, however, that the state space grows exponentially with the number of spatial paths $N^{2}$, inducing a high computational complexity in simulations.  To alleviate the problem, we consider next an approximation for \emph{method~$:1$} that reduces the state space significantly.

In what we call \emph{method~$:2$}, the set of sub-states $\mathcal{S}$ is divided into $K$ disjoint subsets $\{\mathcal{S}_{i} \mid i = 1,\ldots,K\}$.  Each sub-state $\mathbf{s}\in\mathcal{S}_{i}$  corresponds to a variance matrix $\mathbf{V}$ that provides $\rankH = i-1$ spatial DOF.  This leads to a new Markov chain with
\emph{states} $i=1,\ldots,K=N+1$, providing a significant reduction in the state space compared to \emph{method~$:1$}.
%
We remark that \emph{method~$:2$} was used to our knowledge first time in this context in \cite{MIMO:QoS:FSM:Kashif}, where it was found to give accurate results via extensive numerical simulations. In this paper a theoretical justification for \emph{method~$:2$} is provided by the following lemma.

%
%
%
%
%
\begin{lemma}
\label{lemma:cap_approx_highsnr_2}
Let $\mathbf{V}$ be an arbitrary $N \times N$ binary variance matrix for which the degrees of freedom are given by
\begin{equation}
	\rankH(\mathbf{V}) = i - 1 = \mathrm{rank}(\widetilde{\mathbf{H}}), \qquad i = 2,3,\ldots,N+1 ,
\end{equation}
with probability one  when the power constraint~\eqref{E:normalization_v} is preserved.
Then, for sufficiently large $\rankH(\mathbf{V})$, the ergodic capacity at high SNR is given by
\begin{equation}\label{E:HighDOF_HighCap}
\overline{C}(\mathbf{V}) = \rankH(\mathbf{V}) \left[\log_{2} \left(\frac{\rho N}{\rankH(\mathbf{V})}\right) + c'(\mathbf{V})\right],
\end{equation}
where $c'(\mathbf{V})\in O(1)$ is a constant that does not depend on SNR, and
$\rankH(\mathbf{V})$ corresponds to the degrees of freedom, or multiplexing gain, of the virtual channel $\widetilde{\mathbf{H}}$ whose second order statistics are given by the variance matrix $\mathbf{V}$.
\end{lemma}
\begin{IEEEproof}
The proof is deferred to the Appendix.
\end{IEEEproof}
\begin{figure}[t]
\centering
\includegraphics[width=0.9\columnwidth]{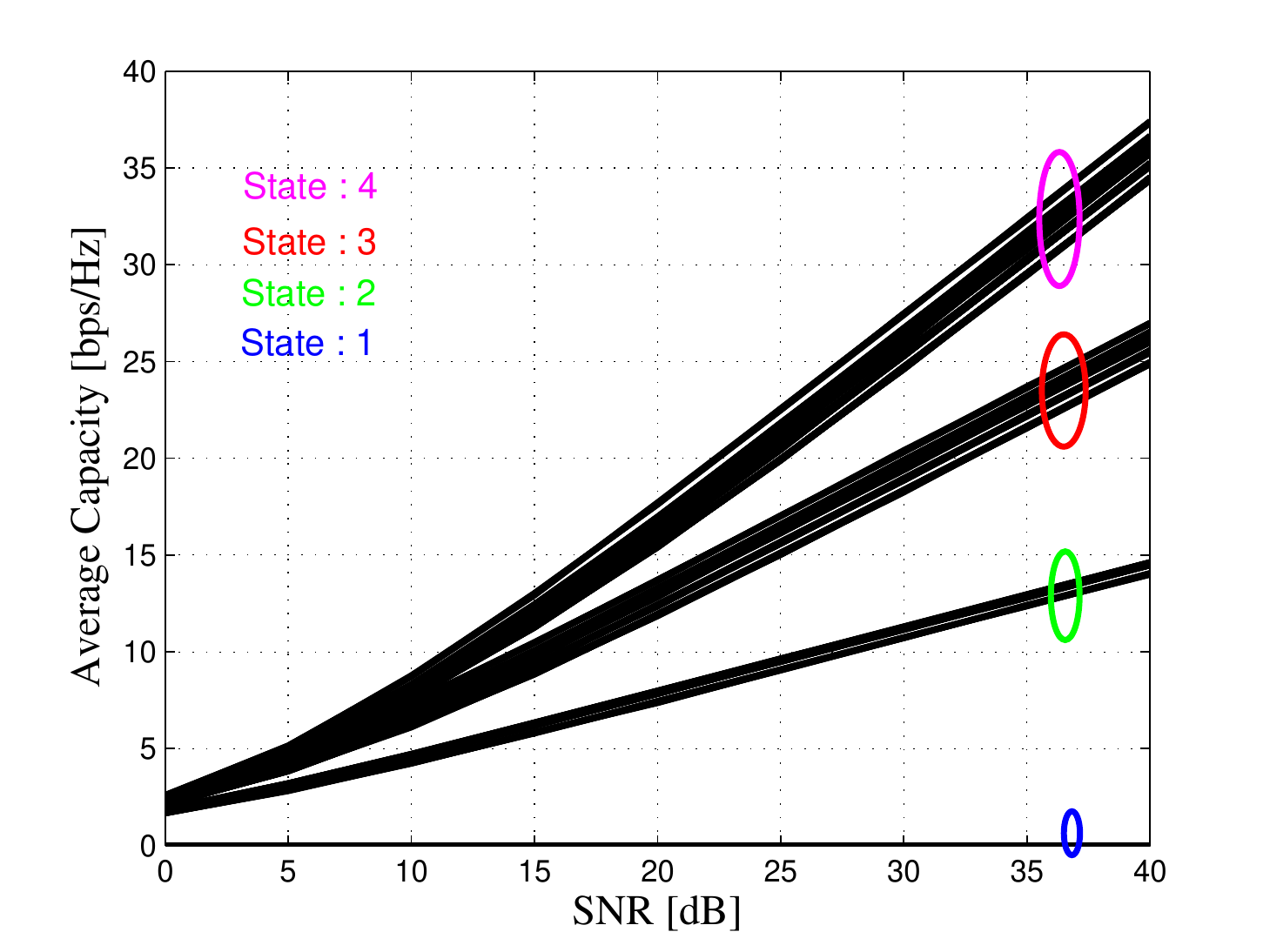}
\caption{States of the Markov Chain for $N = 3$}
\label{F:cap_substates}
\end{figure}
\begin{remark}
	\label{remark:cap_approx_highsnr}
	Lemma~\ref{lemma:cap_approx_highsnr_2} is an extension of the well known high SNR capacity result \cite{MIMO:Foschini:98:limits} where $v_{m,n}=1 \; \forall m,n$, to the case of arbitrary $v_{m,n}\in\{0,v_{\mathbf{s}}\}, \mathbf{s}\in\mathcal{S}$.  As a consequence, for the state $i$ in \emph{method~$:2$}, the ergodic capacity of the system grows  $\propto (i-1) \log_{2} (\rho)$, while the differences between the sub-states are  vertical shifts that become negligible as $\rho\to\infty$.
\end{remark}
%

Fig.~\ref{F:cap_substates} depicts the ergodic capacity of a $3\times3$ MIMO system and shows that the conclusions of the Lemma~\ref{lemma:cap_approx_highsnr_2} are accurate also for small $r$. As noted in Remark~\ref{remark:cap_approx_highsnr}, the difference between the sub-states manifests as a constant vertical shift at high SNR.
The thick lines in Fig.~\ref{F:cap_substates} represent in fact bundles of sub-states.

\begin{defn}

Let $\mathcal{S}_i$ be the set of sub-states and $\barZi$ its cardinality for the state $i=1,\ldots,K$ in \emph{method~$:2$}.  We define the \emph{transmission rate}~$\rate_i$ in state~$i$ as
\begin{equation}
\rate_i=\frac {1}{ \barZi}  \sum_{ \substack {\mathbf{s} \in \mathcal{S}_i } } \overline {C}_{\mathbf{s}} ~, \qquad \forall i ={1,\ldots,K} ,
\label{eq:r_i}
\end{equation}
where $\overline {C}_{\mathbf{s}}$ is the average capacity of the sub-state~$\mathbf{s}\in\mathcal{S}_i$ given by~\eqref{E:MIMOCapFormulalogDetH}.
The set of all rates $\{\rate_i \mid i = 1,\ldots,K\}$ are collected in the diagonal \emph{rate matrix} $\mathbf{R}(\theta)=\mathrm{diag}(e^{\theta \rate_1},\dots,e^{\theta \rate_K})$.
%
\end{defn}

\begin{defn}

Let $ E_{i,j} $ be the event that corresponds to the transition of the Markov chain from state $i$ to state $j$ and let $q_{i,j}$ be the corresponding transition probability 
\begin{equation}\label{E:Qmatrix}
q_{i,j} = \sum_{ \substack {\mathbf{s} \in \mathcal{S}_i } } \pi_{\mathbf{s}}  \cdot  \Pr \left( E_{i,j}\mid \mathbf{s} \right ), \qquad \forall i,j=1,2\ldots K \enspace,
\end{equation}
where the normalized steady state probability~$\pi_{\mathbf{s}}$ of being in sub-state~$\mathbf{s}$ is given as
\begin{equation}\label{E:steady_Prob_substate_unnorm}
	\pi_{\mathbf{s}} = \frac{	\pi_{\mathbf{s}}' }  { \sum_{ \substack {\mathbf{s} \in \mathcal{S}_i } } \pi_{\mathbf{s}}' }\enspace,
\end{equation}
and
\begin{equation}\label{E:steady_Prob_substate}
	\pi_{\mathbf{s}}' = \prod^{N}_{m,n=1} \pi_{\mathbf{s}}^{m,n},  \qquad \pi_{\mathbf{s}}^{m,n} \in \{\kappa,1-\kappa\}\enspace,
\end{equation}
while for all $i,j=1,2\ldots K$
\begin{equation}\label{E:QProb_Cond}
\Pr \left (E_{i,j}\mid \mathbf{s} \in \mathcal{S}_i \right)  = \sum_{\mathbf{s}' \in \mathcal{S}_j }
\prod_{ \substack{m,n=1\\x = S(\widetilde{h}_{m,n})\ltimes \mathbf{s}\\ y = S(\widetilde{h}_{m,n}) \ltimes \mathbf{s}'} }^{N}
p_{x y} \enspace, 
\end{equation}
where $\ltimes$ denotes the element of the vector
and
$p_{xy} \in \{ p_{\mathsf{gb}},1-p_{\mathsf{gb}}, p_{\mathsf{bg}}, 1 -p_{\mathsf{bg}}  \}.$ 
Note that the entries $q_{i,j}$ in \eqref{E:Qmatrix} comprise the elements of the transition probability matrix $\mathbf{Q}$.
The steady state vector $ \boldsymbol{\pi}~=~\left[ \pi_1 \; \pi_2 \;\cdots\,\;\pi_K \right] $ can then be obtained by solving $\boldsymbol{\pi}~=~\boldsymbol{\pi}  \mathbf{Q}$, and we define the first-order capacity~$C^1$ as
\begin{equation}
C^1=\sum_{i=1}^K \rate_i\pi_i .
\label{E:first_order_cap}
\end{equation}
%
%

\end{defn}

\begin{example} {\bfseries ($\boldsymbol{2\times2}$ MIMO system):}
	Let us consider a MIMO system that has $2 \times 2$ antenna setup and uses spatial multiplexing at the transmitter.
	The correlation structure of the channel matrix $\mathbf{H}$ is induced by the virtual domain variance matrix $\mathbf{V}$ and can be obtained from \cite[Eq.~(9)]{Raghavan:2010:WKM}, as explained earlier.  Here, however, we are not interested in the specific forms of spatial correlation, but
	rather describe how to model this simple MIMO system 	
	by a three state Markov chain for which the states $1, 2$ and $3$ correspond to a transmission scenario where the channel has DOF of $0, 1$ and $2$, respectively.

Let us write the sub-states in \emph{method~$:1$} explicitly in the form
$\mathbf{s} = \big[S(\widetilde{h}_{1,1}) \; S(\widetilde{h}_{1,2}) \; S(\widetilde{h}_{2,1}) \; S(\widetilde{h}_{2,2})\big]$.  The \emph{method~$:2$} groups the sub-states $\mathbf{s}\in \mathcal{S}$ of the original Markov chain as:
\begin {itemize}
\item State $1$: $[\mathsf{bbbb}]$.
\item State $2$: $[\mathsf{bbgg}]$, $[\mathsf{bgbg}]$, $[\mathsf{gbbb}]$, $[\mathsf{bgbb}]$, $[\mathsf{bbgb}]$, $[\mathsf{bbbg}]$, $[\mathsf{gbgb}]$, $[\mathsf{ggbb}]$.
\item State $3$: $[\mathsf{bggg}]$, $[\mathsf{gbbg}]$, $[\mathsf{bggb}]$, $[\mathsf{gbgg}]$, $[\mathsf{ggbg}]$, $[\mathsf{gggb}]$, $[\mathsf{gggg}]$.
\end {itemize}
As we saw in Lemma~\ref{lemma:cap_approx_highsnr_2} and Fig.~\ref{F:cap_substates}, the rate at the physical layer is approximately equal for all sub-states that fall within the same DOF when $\mathbf{V}$ is a binary matrix.
The transition matrix $\mathbf{Q}$ of this three state process can be calculated using \eqref{E:Qmatrix}~--~\eqref{E:QProb_Cond}.
The steady state vector follows accordingly and is given by
\begin{IEEEeqnarray*}{rCl}
\boldsymbol{\pi} &=& [\kappa^4 \quad 4\kappa^2(1-\kappa)^2 + 4\kappa^3(1-\kappa)\\
&& \;(1-\kappa)^4 + 4\kappa(1-\kappa)^3 + 2\kappa^2(1-\kappa)^2 ]
\in \mathbb{R}^{1\times3}
\enspace.
\end{IEEEeqnarray*}
The corresponding rate matrix reads
\[\mathbf{R}(\theta)=\mbox{diag}(1,e^{\theta \rate_2},e^{\theta \rate_3}),\]
where we assume that there is no workload processed in state~$1$.

\end{example}

\subsection{Delay Constrained Throughput}
We are now in a position to present the main result of this paper.
The moment generating functions (MGF) of a stationary process $X(t)$ is given by $\mathsf{M}_X(\theta,t) = \Ex \left[e^{\theta X(t)}\right]$, where $\Ex \left[\cdot\right]$ is the expectation.
In the sequel, we denote
$\widehat{\mathsf{M}}_X(\theta,t) = \mathsf{M}_X(-\theta,t)$ for parameter $\theta > 0$, $t \geq 0$.
The discrete time arrivals and service of the channel are assumed to be statistically independent and stationary random processes and are given by the cumulative processes $A(0,t)$ and $S(s,t)$ respectively.
For all real $\theta$, the arrivals and the service random process posses the MGFs $\mathsf{M}_A(\theta,t)$ and $\widehat{\mathsf{M}}_S(\theta,t)$ respectively.
\begin{lemma}\label{Lema:MGF_s}
The MGF of the random process $S(t)$ described by a homogeneous Markov chain, with transition matrix~$\mathbf{Q}$ and stationary state distribution vector~$\boldsymbol{\pi}$, is given for $\theta > 0$ and $t \geq 0$ by
\begin{equation}
\label{E:MS}
\widehat{\mathsf{M}}_S(\theta,t) = \boldsymbol{\pi} (\mathbf{R}(-\theta)\mathbf{Q})^{t-1}\mathbf{R}(-\theta)\mathbf{1} ,
\end{equation}
where $\mathbf{1}$ is column vector of ones while
\begin{equation}
	 \mathbf{R}(\theta) = \mathrm{diag}(e^{\theta \rate_1},\dots,e^{\theta \rate_K}) .
\end{equation}
\end{lemma}
\begin{IEEEproof}
We refer the interested reader to \cite{NetCal:Chang00:PerGuaran:Bk} for the proof.
\end{IEEEproof}
We denote the traffic arrival rate by~$\lambda$ while $d^{\varepsilon}_{\lambda}$ represents a bound with violation probability $\varepsilon$ on the delay.
We can only provide a delay guarantee $d^G$ if $d^{\varepsilon}_{\lambda}\!\leq\!d^G$.
\begin{proposition}
The delay constrained throughput~$\lambda_d$ of a correlated MIMO system under a delay guarantee~$d^G$ is given as
\begin{equation}
\label{E:th}
\lambda_d =\mathrm{max}~\{~\lambda ~ \vert ~ d^{\varepsilon}_{\lambda}\leq d^G \}  ,
\end{equation}
where $d^{\varepsilon}_{\lambda}$ assuming FIFO scheduling is given as
\begin{equation}
\label{E:DelayBoundtwice}
\underset{\,}{d^{\varepsilon}_{\lambda}\!=\! \inf_{\theta > 0} \! \left [ \! \mathrm{inf} \! \left [ \! \tau :  \frac{1}{\theta}  \left ( \! \ln \! \sum_{s=\tau}^{\infty} \! \mathsf{M}_A(\theta,s- \tau) \widehat{\mathsf{M}}_S(\theta,s) \! - \! \ln \varepsilon \right ) \leq 0 \! \right ] \! \right ]  ,}
\end{equation}
and $\widehat{\mathsf{M}}_S(\theta,s)$ is given by \eqref{E:MS}.
\end{proposition}
\begin{IEEEproof}
We refer the interested reader to \cite{NetCal:Chang00:PerGuaran:Bk} for the proof of \eqref{E:DelayBoundtwice}.
\end{IEEEproof}
\begin{remark}
    \label{Rem:Dbound}
The delay bound~\eqref{E:DelayBoundtwice} is calculated using stochastic NetCal approach based on MGF~\cite{NetCal:Fidler2006:EoEProbabNetCalWithMGF}.
MGF for a variety of arrival models is available in the literature~\cite{NetCal:Chang00:PerGuaran:Bk} but finding the MGF of the service process~$\widehat{\mathsf{M}}_S(\theta,t)$ is a challenging task.
We address this challenge by making use of Lemma~\ref{Lema:MGF_s} to calculate $\widehat{\mathsf{M}}_S(\theta,t).$
Having obtained $\widehat{\mathsf{M}}_S(\theta,t)$, a stochastic bound on the delay $d^{\varepsilon}_{\lambda}$ in \eqref{E:DelayBoundtwice} can be obtained using Chernoff's bound, Boole's inequality and applying the technique in \cite{NetCal:Fidler2006:EoEProbabNetCalWithMGF}.

\end{remark}

We next use the result in this section to calculate the delay constrained throughput of a $N \times N$ MIMO channel which is parameterized according to IEEE $802.11$n.
\section{Application and Numerical Results}
\label{Sec:NumericalResults}
We present numerical results for the delay constrained throughput of a MIMO system which is parameterized according to 802.11n using $40$ MHz channel~\cite{MIMO:80211n:standard}.
We choose a base time unit of $31\mu s$ for the discrete time model which is approximately the time needed to transmit a data block of $2312$ bytes over a $600$ Mbps channel.
The rate matrix~$\mathbf{R}(\theta)$ is obtained through an extensive simulation as described in Section~\ref{subsec:StateAggBasedonDOF}.
We omit showing confidence intervals as these turn out to be very small.
We fix the SNR to $25$dB for all $N$ and set the rate of the highest state~$r_K$ corresponding to this fixed SNR.
The rate of the rest of the states follows accordingly corresponding to the different channel matrixes.
The infinite sum in the delay bound formula~\eqref{E:DelayBoundtwice} is computed for the first $4000$ units of time.
For the GE model, we set the transition probability from $\mathsf{bad}$ to $\mathsf{good}$ $p_{\mathsf{bg}} = 0.1$ and the probability from $\mathsf{good}$ to $\mathsf{bad}$ $p_{\mathsf{gb}} = 0.01$ similar to \cite{NetCal:Fidler06:MGFfadingChannel}.

The framework discussed earlier enables the derivation of delay constrained throughput of a various number of traffic sources whenever the MGF exists. A collection of such MGFs for sources of different statistical properties can be found in \cite{EB:Kelly96}.
In this work we use a periodic source for the discrete time model that generates arrival traffic.
Such a traffic source with period $\tau$ produces $\sigma$ units of workload (data blocks) at times $\left\{U \tau + n \tau, n = 0,1,\ldots\right\}$ where $U$ is the starting time which is uniformly distributed  in the interval $[0,1]$. The MGF of such a source is known (see e.g.~\cite{EB:Kelly96}) as
\begin{equation}
	\mathsf{M}_{A}(\theta,t) = e^{\theta \sigma \left\lfloor \frac{t}{\tau}\right\rfloor} \left(1+\left(\frac{t}{\tau} - \left\lfloor \frac{t}{\tau}\right\rfloor\right)\left(e^{\theta \sigma} - 1\right)\right) ,	
\end{equation}
for $t \geq 0$ and $\theta > 0$.
We parameterize the source such that its period $\tau$ is 10 time units and set the number of generated data blocks $\sigma$ to achieve a given arrival rate $\lambda$.
\begin{figure}[t]
	\centering		 \includegraphics[width=0.90\columnwidth]{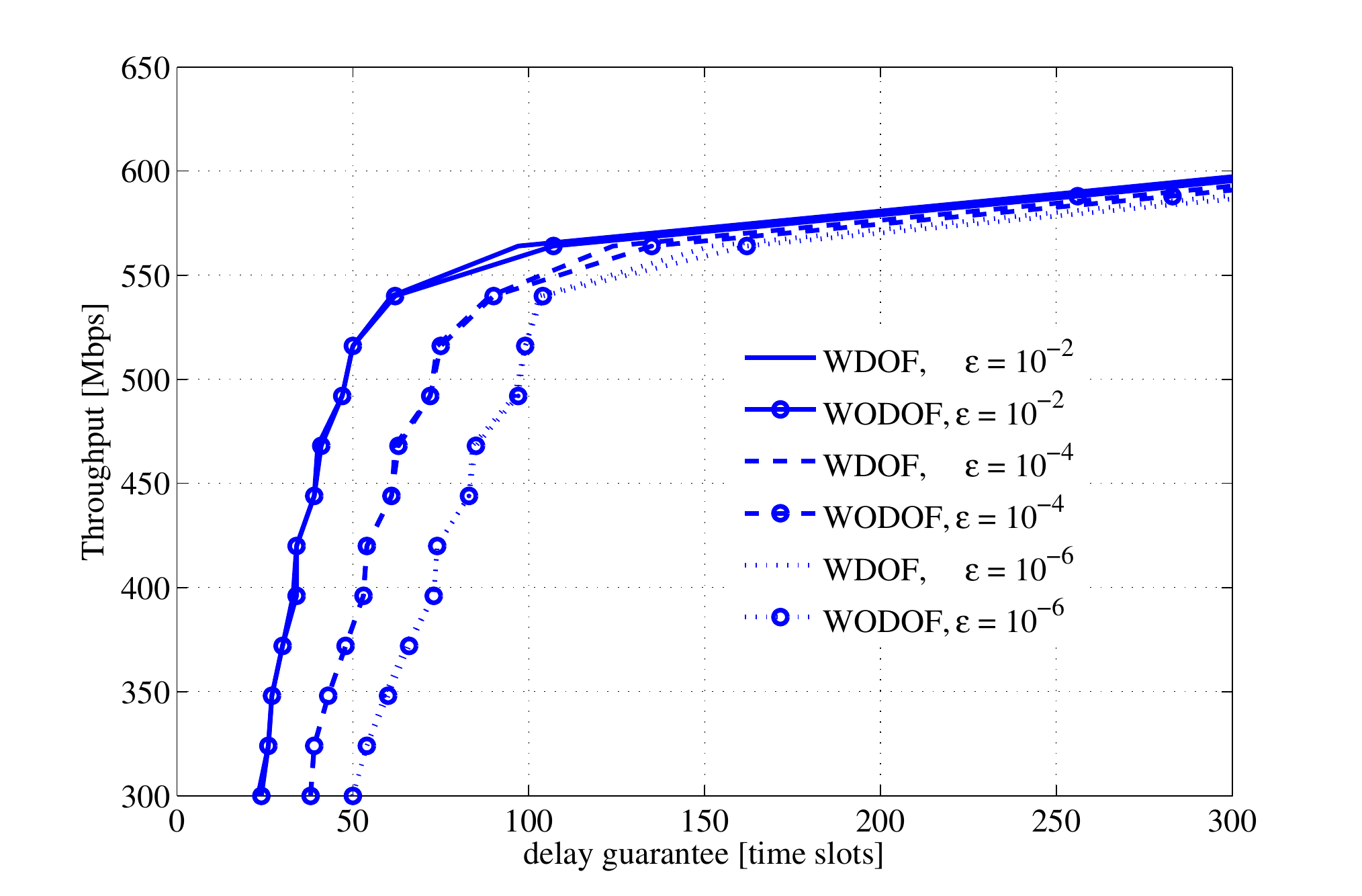}
	\caption{Delay constrained throughput for different violation probabilities~$\varepsilon$ using \emph{method~$:1$}(WODOF) and \emph{method~$:2$}(WDOF) }
	\label{fig:thVSdb_N2}
\end{figure}

Fig.~\ref{fig:thVSdb_N2} depicts throughput as a function of delay guarantee~$d^{G}$ for different delay bound violation probabilities $\varepsilon$ using \emph{method~$:1$} without DOF aggregation (WODOF) and \emph{method~$:2$} with DOF aggregation (WDOF) for $N = 2$ antennas.
It shows that the \emph{method~$:2$} of aggregating the states based on DOF is reasonably accurate.
For the results to follow, we use \emph{method~$:2$} of DOF based aggregation only.
Furthermore, one can observe the decrease in throughput as the violation probability gets tighter.
\begin{figure}[t]
	\centering		 \includegraphics[width=0.90\columnwidth]{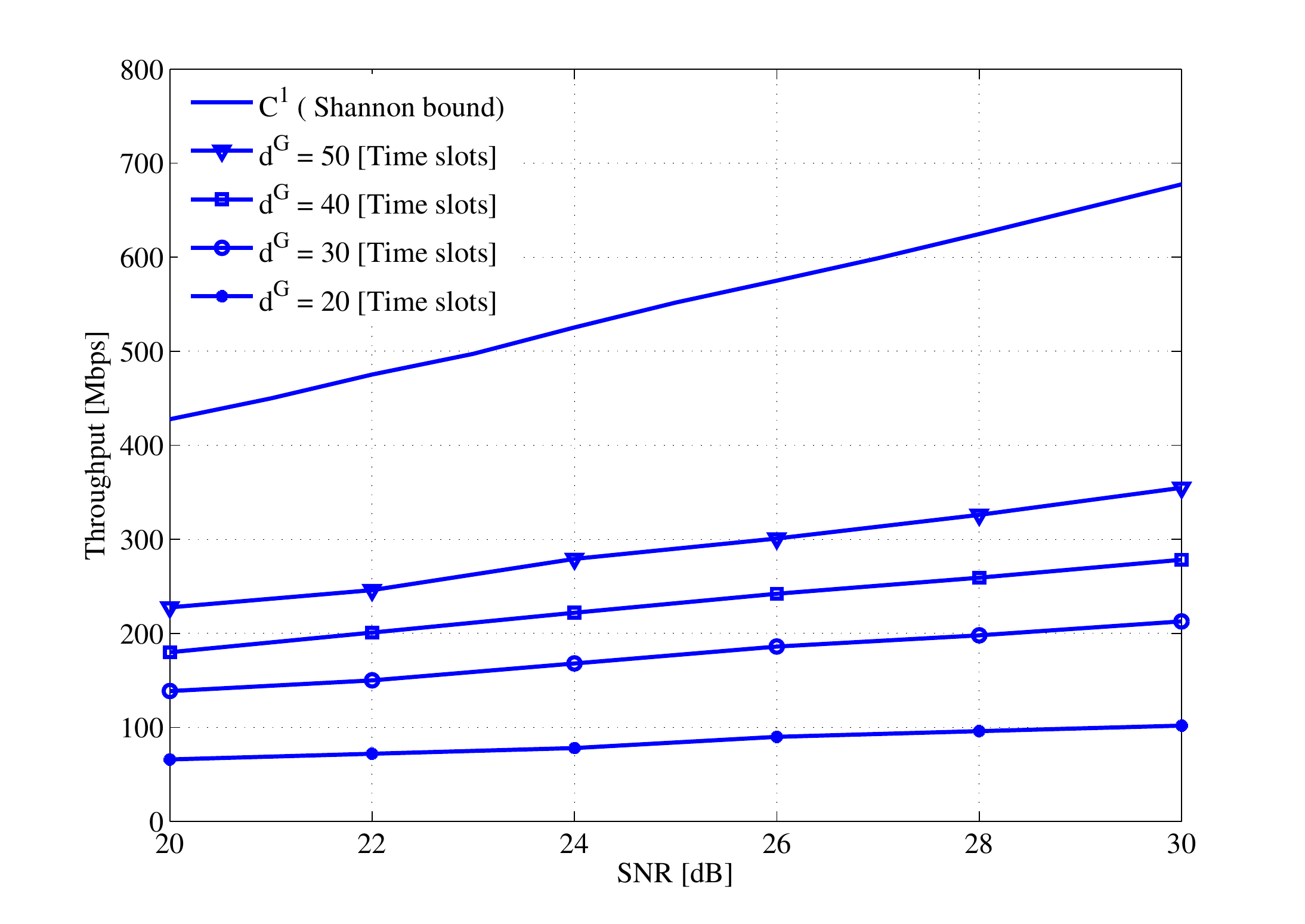}
	\caption{Delay constrained throughput for different SNR's}
	\label{fig:thVSSNR_N2}
\end{figure}

It is to be noted that Shannon limit is the upper bound to the delay constrained throughput as can be seen in Fig.~\ref{fig:thVSSNR_N2} for $N = 2$, where we fix $\varepsilon = 10^{-6}$. Also note that the improvement in throughput is less as we go to the higher values of delay guarantee.

Tab.~\ref{tab:fading_meancap} verifies the known notion that higher antenna configuration provides a higher capacity.
It is to be noted that the ratio of the first-order capacities $C^1$ of $N = 3$ to $N = 2$ stays unchanged over different fading speeds.
\begin{table}[tb]
\begin{center}
\caption{First-order capacity for different \# of antennas}
\label{tab:fading_meancap}
\begin{tabular}{|c||c|} \hline
$N$ & $C^1$ [bps/Hz] \\ \hline
2 & 14.1 \\ \hline
3 & 20.82 \\ \hline
\end{tabular}
\end{center}
\end{table}
\begin{figure}[t]
	\centering		 \includegraphics[width=0.90\columnwidth]{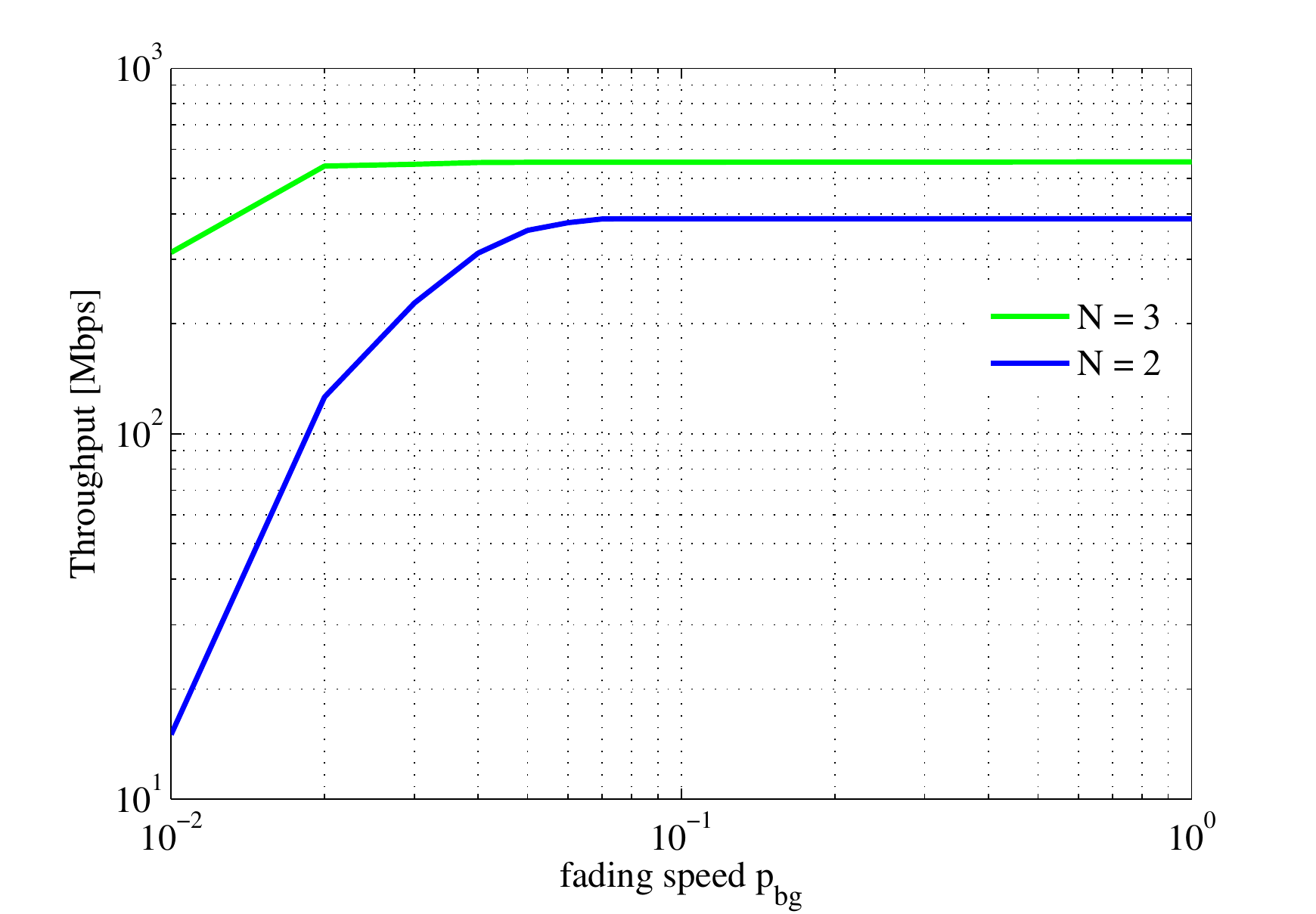}
	\caption{Throughput as a function of fading speed}
	\label{fig:thVSfad}
\end{figure}

We next study the effect of fading speed on the throughput for different number of antennas ($N = 2$, $3$) in Fig.~\ref{fig:thVSfad} on loglog scale where the delay guarantee is fixed to 30 time slots and $\varepsilon = 10^{-3}$.
We fix the steady-state error rate $\kappa$ for each spatial path given by \eqref{eq:block_error_prob} to $0.1$. We control the fading speed, under fixed $\kappa$, by the mean time for the Markov chain of the path to change states twice.
To this end we variate $p_{\mathsf{bg}}$ such that a small value corresponds to slow fading and vice versa.
The notable thing is that for slow fading the difference in throughput between $N = 2$ and $N = 3$ is large and for higher fading speeds, which corresponds to a channel nearly without memory, the profit from extra spatial paths diminishes.
As the first order capacity $C^1$ given by \eqref{E:first_order_cap} and shown in Tab.~\ref{tab:fading_meancap} stays constant over the range of fading speeds, which is implied by the fixed $\kappa$, the difference in throughput might be due to the higher order statistics of capacity.
Furthermore, the impact of statistical multiplexing\footnote{Statistical multiplexing, which is a flow level phenomenon, should not to be confused with spatial multiplexing of the MIMO system.} between the spatial paths, which is higher for $N = 3$, is apparent in the slow fading regime where the Markov chain changes states less frequently.
\begin{figure}[t]
	\centering
		 \includegraphics[width=0.90\columnwidth]{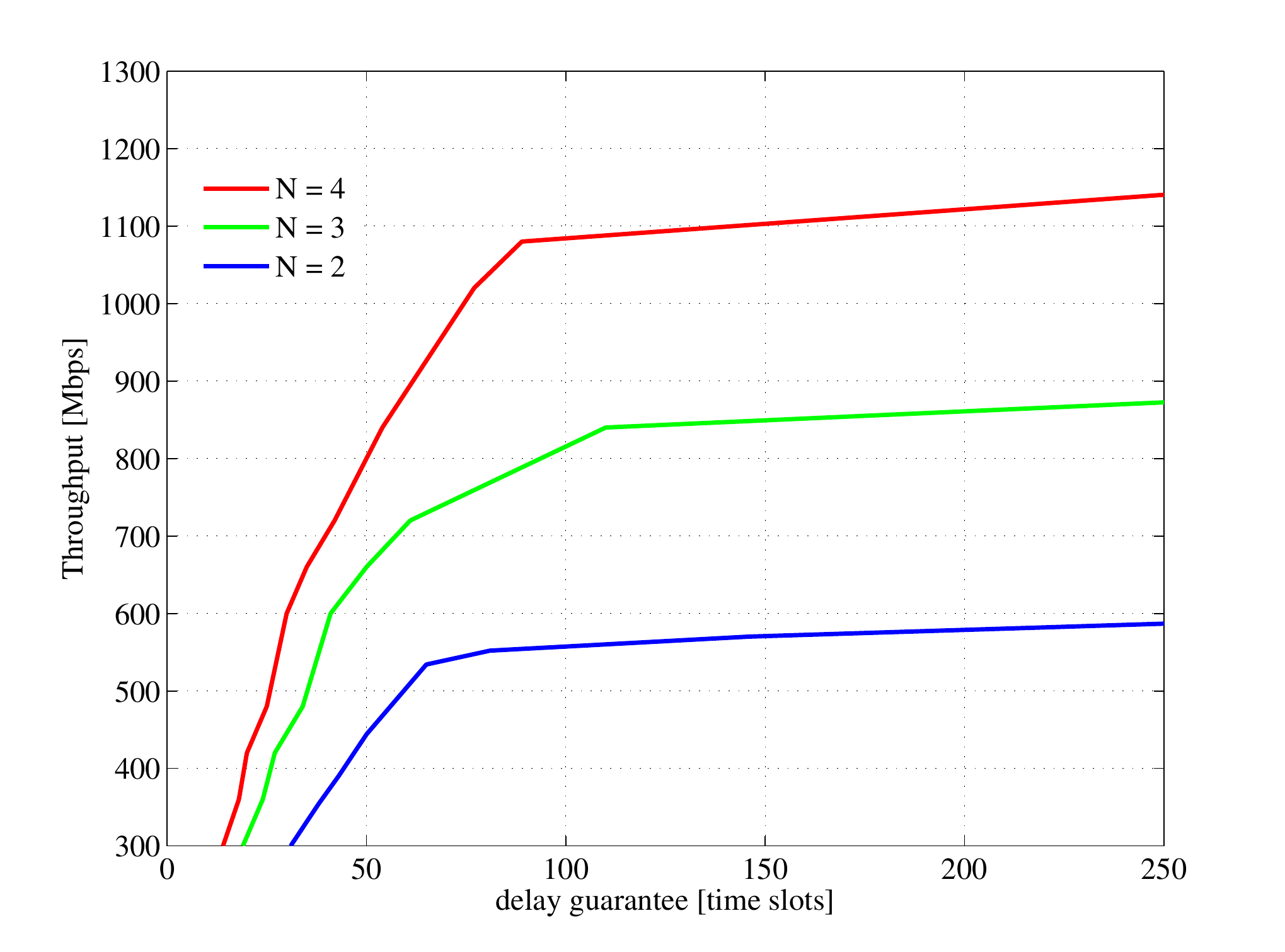}
	\caption{Throughput as a function of different \# of antennas $N$}
	\label{fig:thVSdb_diff_nr_antennas}
\end{figure}

It is well known~\cite{MIMO:Teletar99:CapacityOfMultiAntennaGaus} that an increase in the number of antennas leads in the best case to a linear increase in channel capacity.
This increase capacity manifests itself in the increase of the delay constrained throughput for a fixed scenario conditions.
This can be seen in Fig. \ref{fig:thVSdb_diff_nr_antennas} where $\varepsilon = 10^{-3}$.
It saturates after a given delay guarantee owing to the well understood concept that traffic arrival rate should never be greater than the capacity to ensure system stability.
%
%
\section{Conclusions and Future Work}
In this paper we formulate a method to find the delay constrained throughput of a correlated, spatially multiplexed MIMO system in which the MIMO wireless channel is modeled as a Markov chain.
The states of the Markov chain are decided on the degree of freedom for which we provide a rigorous mathematical proof.
We parameterize the model to the application of IEEE 802.11n.
We present numerical results where we quantify the impact of increasing the delay guarantee and the number of antennas on the delay constrained throughput for varying signal strength and fading speed.

This work finds it application in performance evaluation of wireless networks where we are interested in the maximum attainable throughput for a given delay guarantee and resources of the network.
While in this work, we only used periodic source to illustrate the results, the same methodology can be applied to any traffic source with known MGF.
The formulation can also be extended to multihop and multi-user scenarios and it is left as our future work.
\section*{Acknowledgment}
The first author would like to thank Amr Rizk for the useful discussions.

\section*{Appendix}
\label{Ap:Aggregation}

Here we sketch the proof of Lemma~\ref{lemma:cap_approx_highsnr_2} introduced in Section~\ref{Sec:MarkovModleingCorMIMO}.  The result is obtained in the large system limit where the (effective) number of antennas on both sides grow without bound.
We consider only $N \times N$  Gaussian MIMO channels that are described in the virtual domain by a binary variance matrix $\mathbf{V}$ with $\rankH$ non-zero rows and columns.  The derivation for general $M \times N$ MIMO channel as well as the case of unequal number of non-zero rows and columns follows by the same procedures but is omitted in the present paper due to space constraints.
We also remark that our result is akin to the one proposed in \cite{Raghavan:2010:WKM}, but the proof is slightly different due to the special structure of the matrix $\mathbf{V}$ considered in this paper.

%
\begin{lemma}
	\label{lemma:cap_approx_highsnr}
	Let $\rankH = \mathrm{rank}(\widetilde{\mathbf{H}})$ with probability one, and let $\mathbf{V}$ be such an $N\times N$ binary variance matrix that by deleting the all-zero rows and columns from it we get a modified $\rankH \times \rankH$ variance matrix $\modV$.  Denote the number of non-zero elements in the row $m=1,2,\ldots,\rankH$ of $\modV$ by
	$D_{m}$,
	let $D = \sum_{m=1}^{\rankH}D_{m}$ be the total number of non-zero elements in $\modV$ $(\text{and } \mathbf{V})$ and
	write $\mathcal{D}=\{ D_{m} : m = 1,\ldots,\rankH\}$.
	Then, for fixed $\mathbf{V}$, the ergodic capacity 
reads
	\begin{IEEEeqnarray}{rCl}
		\frac{\overline{C}(\mathcal{D})}{\rankH} &=&
		\log_{2} (\rho N)
				- \frac{1+\gamma}{\ln (2)}		\IEEEnonumber\\
				&& \quad +
				\underbrace{
				\frac{1}{\rankH\ln(2)} \sum_{m=1}^{\rankH}  H_{D_{m}-1} - \log_{2} (D)}_{\triangleq c(\mathcal{D})} , \IEEEeqnarraynumspace
		\label{eq:cap_approx_highsnr_largemimo_1} 	
	\end{IEEEeqnarray}
	where $\rankH\to\infty$, $\gamma = \lim_{n\to\infty} \big( H_{n} - \ln(n) \big) =
	0.57721566 \ldots$ is the Euler-Mascheroni constant and
		$H_{n} = \sum_{p=1}^{n} 1/p$, the $n$th harmonic number.
\end{lemma}

\begin{IEEEproof}
Let $\modH\in\mathbb{C}^{\rankH \times \rankH}$ be a modified Gaussian virtual MIMO channel that is
obtained from $\widetilde{\mathbf{H}}$ by removing the all-zero rows and columns.
The non-zero elements $\{\hat{h}_{m,n}>0: m,n = 1,\ldots,\rankH\}$ of $\modH$ are independent by assumption \cite{MIMO:Veeravalli2005:CorrelatedMIMO:Variance} and their variances are given by the corresponding elements of $\modV$.  The ergodic capacity per DOF at high SNR, given $\modV$, can then be written as \cite[Eq.(78)]{Raghavan:2010:WKM}
	\begin{equation}
		\label{eq:exp_logdet_1}
		\underset{\,}{\frac{\overline{C}(\modV)}{\rankH} =  \frac{1}{\rankH} \Ex \log_{2} \big [\det \big (\modH \modH^{\dag}   \big )     \big ]
		+\log_{2} \left(\frac {\rho}{N}\right) + O\left(\frac{1}{\rankH\rho}\right).}
	\end{equation}
	Note that the last term in \eqref{eq:exp_logdet_1} vanishes in the high-SNR limit $\rho\to\infty$.  The second term
	depends on the total number of transmit antennas $N$ since the transmitter has no (statistical) channel information and uses therefore uniform power allocation over all antennas.  By \cite[Lemma~5]{Raghavan:2010:WKM},
	\begin{IEEEeqnarray}{l}
		\label{eq:exp_logdet_2}
		\lim_{\rankH\to\infty}\frac{1}{\rankH}
		\Ex \log_{2} \big [\det \big (\modH\modH^{\dag}   \big )     \big ]
		= \lim_{\rankH\to\infty}\frac{1}{\rankH}\bigg\{\log_{2} \bigg(\frac{N^2}{D}\bigg) \IEEEeqnarraynumspace\IEEEnonumber\\[1ex]
		\qquad+ \frac{1}{\rankH\ln(2)} \sum_{m=1}^{\rankH} \Ex \ln
		\bigg( \frac{m}{\rankH} \underbrace{\sum_{n=1}^{\rankH} \sqrt{\frac{D}{N^{2}}} |\hat{h}_{m,n}|^2}_{\sim \chi^{2}_{2 D_{m}}}
		\bigg) \bigg\} \enspace,  \IEEEeqnarraynumspace  \\[-2em] \IEEEnonumber
	\end{IEEEeqnarray}
	where $\chi^{2}_{2 D_{m}}$ is a chi-square random variable with $2 D_{m}$ degrees of freedom.
	The expectation in \eqref{eq:exp_logdet_2} has a closed form solution
	(see e.g., \cite[Lemma~10.1]{Lapidoth:2003:CapacityDuality})
	\begin{equation}
		\label{eq:exp_of_logchisq}
		\Ex \ln (\chi^{2}_{2 D_{m}}) =  H_{D_{m}-1} - \gamma \enspace,
	\end{equation}
	so that plugging \eqref{eq:exp_logdet_2}~--~\eqref{eq:exp_of_logchisq} to \eqref{eq:exp_logdet_1} and simplifying yields at high SNR
	\begin{IEEEeqnarray}{l}
		\lim_{\rankH\to\infty}\frac{\overline{C}(\mathcal{D})}{\rankH} \IEEEnonumber\\
		\quad= \;
		\lim_{\rankH\to\infty}\left\{
		\log_{2} \left(\frac {\rho N}{ D}\right) + 		
		\frac{1}{\rankH\ln(2)} \sum_{m=1}^{\rankH} H_{D_{m}-1}
		\right\}
		-\frac{1+\gamma}{\ln (2)} \enspace, \IEEEnonumber\\
		\label{eq:exp_logdet_3}
	\end{IEEEeqnarray}	
	completing the proof. 
\end{IEEEproof}

\begin{remark}
	For large degrees of freedom $\rankH$, and a square (modified) variance matrix $\modV$, the ergodic capacity~\eqref{E:MIMOCapFormulalogDetH} depends on  $\mathbf{V}$ only through $c(\mathcal{D})$ in \eqref{eq:cap_approx_highsnr_largemimo_1}.
The exact structure of $\mathbf{V}$ is thus not important, just the number of non-zero paths between the receive antennas and the transmit antennas.  Furthermore,
for fixed $\rankH$ and $\mathcal{D}$, the maximum capacity difference between the sub-states can be explicitly calculated from $c(\mathcal{D})$.  In the following, however, we do not consider the exact calculation of $c(\mathcal{D})$, but rather study its behavior in the high-SNR and large system limits.  This will provide and expression for capacity that holds for all choices of $\mathcal{D}$ as stated in Lemma~\ref{lemma:cap_approx_highsnr_2}.
\end{remark}

\subsection*{Proof of Lemma~\ref{lemma:cap_approx_highsnr_2}}\label{pf:Lemma_2}
Let us first consider the case when $\mathbf{V}$ is not sparse, that is,
$D_{m} \in O(\rankH)$ or
\begin{IEEEeqnarray}{l}
	\label{eq:cond_Dm_1}
	\lim_{\rankH\to\infty} \frac{D_{m}}{\rankH} = \alpha_{m}, \qquad 0<\alpha_{m}<1, \qquad m = 1,\ldots,\rankH \enspace. \IEEEeqnarraynumspace
\end{IEEEeqnarray}
From Lemma~\ref{lemma:cap_approx_highsnr} and the bound \cite{GammaEulersConstant:2003}
\begin{equation}
	\frac{1}{24(n+1)^{2}} < H_{n} - \big(\ln(n + 1/2) + \gamma\big) < \frac{1}{24n^{2}} \enspace,
\end{equation}
we have
\begin{IEEEeqnarray}{l}
	\label{eq:c_D_infty}
	\lim_{\rankH\to\infty} c(\mathcal{D}) =
	\frac{\gamma}{\ln(2)} \IEEEnonumber\\
	\quad +\lim_{\rankH\to\infty}
			\frac{1}{\rankH \ln(2)}
			\sum_{m=1}^{\rankH} \bigg[\ln \bigg(\frac{D_{m}-1/2}{D}\bigg)
			+ O\bigg(\frac{1}{D^{2}_{m}}\bigg)\bigg], \IEEEeqnarraynumspace
\end{IEEEeqnarray}	
where the last term in \eqref{eq:c_D_infty} vanishes rapidly with $\{D_{m}\}_{m=1}^{\rankH}$.  By definition
\[
\lim_{\rankH\to\infty} \frac{D_{m}-1/2}{D}
= \lim_{\rankH\to\infty}\frac{\alpha_{m}}{\sum_{n=1}^{\rankH} \alpha_{n}} \enspace,
\]
so that
\begin{equation}
	\label{eq:cD_1}
	\lim_{\rankH\to\infty} c(\mathcal{D}) =
	\frac{\gamma}{\ln(2)}+
	\lim_{\rankH\to\infty}
			\frac{1}{\rankH}
			\sum_{m=1}^{\rankH} \log_{2} \bigg(\frac{\alpha_{m}}{\sum_{n=1}^{\rankH} \alpha_{n}}\bigg) \enspace.
\end{equation}
Plugging \eqref{eq:cD_1} to Lemma~\ref{lemma:cap_approx_highsnr} shows that as $\rankH\to\infty$,
\begin{equation}
			\label{eq:cap_approx_highsnr_largemimo_2}
	\frac{\overline{C}(\mathcal{D})}{\rankH} =
	\log_{2} (\rho N)
	-\frac{1}{\ln (2)}
	+\frac{1}{\rankH}\sum_{m=1}^{\rankH} \log_{2} \bigg(\frac{\alpha_{m}}{\sum_{n=1}^{\rankH} \alpha_{n}}\bigg)
	\enspace.
\end{equation}
Thus, for all $\mathbf{V}$ that satisfy \eqref{E:normalization_v} and \eqref{eq:cond_Dm_1},
at high SNR and for large but fixed degrees of freedom $\rankH$, the ergodic capacity is given by
\begin{equation}
	\label{eq:cap_approx_highsnr_largemimo_4}
	\overline{C}(\mathcal{D}) = \rankH \left[\log_{2} \left(\frac {\rho N}{ \rankH}\right) + c'(\mathcal{D}) \right] \enspace,
\end{equation}
where $c'(\mathcal{D})\in O(1)$.

By similar calculus as above, it is not difficult to show that
\eqref{eq:cap_approx_highsnr_largemimo_4} holds also if
the matrix $\mathbf{V}$ is sparse, that is,
$D_{m} \in o(\rankH)$ or
\begin{equation}
	\label{eq:cond_Dm_2}
	\lim_{\rankH\to\infty} \frac{D_{m}}{\rankH} = 0, \qquad \forall m = 1,\ldots,\rankH \enspace,
\end{equation}
where the $D_{m}$'s are finite constants or grow sub-linearly with $r$ (for example, $D_{m} = \sqrt{\rankH}$).
Furthermore, Lemma~\ref{lemma:cap_approx_highsnr} can be modified for non-square $\modV$, and the rest of the proof follows a similar calculus as shown here.  The final result is that \eqref{eq:cap_approx_highsnr_largemimo_4} holds also for such variance matrices, as stated in Lemma~\ref{lemma:cap_approx_highsnr_2}. These cases lead, however, to more cumbersome notation and we have omitted them in the proof due to space constraints.


\bibliographystyle{IEEEtran}
\bibliography{./XBib_Kashif_Amr_Mikko}

\end{document}